\documentclass[prl,twocolumn,groupedaddress,showpacs,nofootinbib]{revtex4}
\usepackage{graphicx}
\usepackage{dcolumn}
\usepackage{bm}
\usepackage{amssymb}
\usepackage{mathrsfs}
\usepackage{amsmath}
\usepackage{epsfig}
\usepackage[dvips]{color}
\usepackage{hhline}

\begin{document}

\title{Universal Seesaw from Left-Right and Peccei-Quinn Symmetry Breaking}

\author{Pei-Hong Gu}
\email{peihong.gu@mpi-hd.mpg.de}

\author{Manfred Lindner}
\email{manfred.lindner@mpi-hd.mpg.de}

\affiliation{Max-Planck-Institut f\"{u}r Kernphysik, Saupfercheckweg
1, 69117 Heidelberg, Germany}

\begin{abstract}

To generate the lepton and quark masses in the left-right symmetric
models, we can consider a universal seesaw scenario by integrating
out heavy fermion singlets which have the Yukawa couplings with the
fermion and Higgs doublets. The universal seesaw scenario can also
accommodate the leptogenesis with Majorana or Dirac neutrinos. We
show the fermion singlets can obtain their heavy masses from certain
global symmetry breaking, which is driven by one complex scalar
singlet or two. The global symmetry can be identified to the
Peccei-Quinn symmetry since it is mediated to the standard model
quarks at tree and/or loop level.

\end{abstract}

\pacs{12.60.Cn, 12.60.Fr, 14.80.Va, 98.80.Cq}

\maketitle

\emph{Introduction}: In the $SU(3)_c^{}\times SU(2)_{L}^{}\times
U(1)_{Y}^{}$ standard model (SM), the charged fermion singlets have
Yukawa interactions with the fermion and Higgs doublets so that the
quarks and the charged leptons can obtain their Dirac masses after
the Higgs doublet develops its vacuum expectation value (VEV).
However, we can not get the neutrino masses in this way because the
right-handed neutrinos are absent in the SM. To naturally generate
the neutrino masses, which are far lower than the charged fermion
masses, we can consider the seesaw \cite{minkowski1977} extension of
the SM by introducing the right-handed neutrino singlets with heavy
Majorana masses \cite{minkowski1977} and/or the left-handed Higgs
triplet(s) with small VEV(s) \cite{mw1980}.

The SM and its seesaw extension can be embedded in the
$SU(3)_c^{}\times SU(2)_{L}^{}\times SU(2)_{R}^{}\times
U(1)_{B-L}^{}$ left-right symmetric models \cite{ps1974}, where the
left(right)-handed fermions are placed in
$SU(2)_{L}^{}[SU(2)_{R}^{}]$ doublets. For the breakdown of
$SU(2)_R^{}\times U(1)_{B-L}^{}$ to $U(1)_{Y}^{}$, the simplest
choice is to introduce a right-handed Higgs doublet and its
left-handed partner. The fermion doublets associated with the
additional fermion singlets can have the Yukawa couplings with the
Higgs doublets \cite{berezhiani1983,cm1987,bms2003,gu2010}.
Subsequently we can integrate out the fermion singlets to generate
the masses of the quarks, the charged leptons, and the neutral
neutrinos. In this scenario, the seesaw is a universal origin of the
fermion masses. The universal seesaw can also accommodate the
leptogenesis \cite{fy1986} mechanism with Majorana \cite{fy1986} or
Dirac \cite{dlrw1999} neutrinos to explain the matter-antimatter
asymmetry in the universe. The key point of the universal seesaw is
the existence of the heavy fermion singlets, including the color
singlets for generating the lepton masses and the color triplets for
generating the quark masses. The heavy masses of the fermion
singlets can be simply input by hand as they are allowed by the
gauge symmetry. A more interesting possibility is that the fermion
singlets obtain their masses through certain spontaneous symmetry
breaking. For example, the original work on the universal seesaw
introduced a new $SU(3)$ gauge symmetry \cite{berezhiani1983}.

In this paper we shall consider a spontaneous Peccei-Quinn (PQ)
symmetry \cite{pq1977} breaking to generate the heavy masses of the
fermion singlets for the universal seesaw. Specifically we shall
impose a global symmetry under which the left- and right-handed
fermion singlets carry equal but opposite charges. Accordingly, the
left- and right-handed fermion and/or Higgs doublets also carry
equal but opposite charges through their Yukawa couplings with the
fermion singlets. In this context, the left-right symmetry should be
the charge-conjugation. The spontaneous symmetry breaking of the
global symmetry is driven by one complex scalar or two. The fermion
singlets can obtain the heavy masses through their Yukawa couplings
with the complex scalar singlet(s). The Nambu-Goldstone boson (NGB),
associated with the global symmetry breaking, can become a pseudo
NGB (pNGB) as it picks up a tiny mass through the color anomaly
\cite{adler1969}. Since the global symmetry is mediated to the SM
quarks at tree and/or loop level, it can be identified with the PQ
symmetry to solve the strong CP problem. Consequently, the pNGB
should be an invisible \cite{kim1979,dfs1981} axion
\cite{pq1977,weinberg1978}.

\emph{The Model}: The Higgs scalars include a left-handed doublet
$\phi_L^{}(\textbf{1},\textbf{2},\textbf{1},-1)$, a right-handed
doublet $\phi_R^{}(\textbf{1},\textbf{1},\textbf{2},-1)$, a real
singlet $\sigma(\textbf{1},\textbf{1},\textbf{1},0)$ and one complex
singlet or two $\xi(\textbf{1},\textbf{1},\textbf{1},0)$. The full
potential is
\begin{eqnarray}
\label{potential}
\hskip -1.1cmV&=&\mu_1^2\sigma^2_{}+\mu_2^2|\xi|^2_{}+\mu^3_{3}(|\phi_L^{}|^2_{}+|\phi_R^{}|^2_{})+\lambda_1^{}\sigma^4_{}+\lambda_2^{}|\xi|^4_{}\nonumber\\
&&
+\lambda_3^{}(|\phi_L^{}|^4_{}+|\phi_R^{}|^2_{})+\lambda_4^{}|\phi_L^{}|^2_{}|\phi_R^{}|^2_{}+\kappa_1^{}\sigma^2_{}|\xi|^2_{}\nonumber\\
&&+\mu_4^{}\sigma(|\phi_L^{}|^2_{}-|\phi_R^{}|^2_{})
+\kappa_2^{}\sigma^2_{}(|\phi_L^{}|^2_{}+|\phi^{}_R|^2_{})\nonumber\\
&&+\kappa_3^{}|\xi|^2_{}(|\phi_L^{}|^2_{}+|\phi^{}_R|^2_{})\,.\phantom{xxx}
\end{eqnarray}
Here we have taken the left-right symmetry to be the
charge-conjugation, which yields
\begin{eqnarray}
\label{lrsymmetry1} \sigma\leftrightarrow-\sigma\,,\quad
\xi\leftrightarrow\xi\,,
\quad\phi_{L}^{}\leftrightarrow\phi_{R}^{\ast}\,.
\end{eqnarray}
Note that the parity-odd singlet $\sigma$ is responsible for the
spontaneous D-parity violation \cite{cmp1984} to guarantee the
different VEVs of the left- and right-handed Higgs doublets
\cite{gu2010}. We may replace the spontaneous D-parity violation by
softly breaking the left-right symmetry. In the fermion sector,
besides the usual quark and lepton doublets,
$q_{L}^{}(\textbf{3},\textbf{2},\textbf{1},\frac{1}{3})$,
$q_{R}^{}(\textbf{3},\textbf{1},\textbf{2},\frac{1}{3})$,
$l_{L}^{}(\textbf{1},\textbf{2},\textbf{1},-1)$, and
$l_{R}^{}(\textbf{1},\textbf{1},\textbf{2},-1)$, there are four
types of fermion singlets,
$D_{L,R}^{}(\textbf{3},\textbf{1},\textbf{1},-\frac{2}{3})$,
$U_{L,R}^{}(\textbf{3},\textbf{1},\textbf{1},\frac{4}{3})$,
$E_{L,R}^{}(\textbf{1},\textbf{1},\textbf{1},-2)$, and
$N_{L,R}^{}(\textbf{1},\textbf{1},\textbf{1},0)$. Under the
left-right symmetry, the fermions transform as,
\begin{eqnarray}
\label{lrsymmetry2}
\begin{array}{ccc}
q_{L}^{}\leftrightarrow q_{R}^{c}\,,&D_{L}^{}\leftrightarrow
D_{R}^{c}\,,&U_{L}^{}\leftrightarrow U_{R}^{c}\,,\\
l_{L}^{}\leftrightarrow l_{R}^{c}\,,&\,E_{L}^{}\leftrightarrow
E_{R}^{c}\,,&N_{L}^{}\leftrightarrow N_{R}^{c}\,.
\end{array}
\end{eqnarray}

We further impose a global symmetry under which the fields carry the
following quantum numbers,
\begin{eqnarray}
\label{pqsymmetry} \begin{array}{ccccc} x_1^{}& \textrm{for}&
D_L^{}\leftrightarrow D_R^c&\textrm{and}&
E_L^{}\leftrightarrow E_R^c\,,\\
x_2^{}&\textrm{for}&U_L^{}\leftrightarrow
U_R^c&\textrm{and}&N_L^{}\leftrightarrow
N_R^c\,,\\
y&\textrm{for}&q_L^{}\leftrightarrow
q_R^c&\textrm{and}&l_L^{}\leftrightarrow l_R^c\,,
\end{array}~
\begin{array}{ccc}2x_1^{}&\textrm{for}&\xi_1^{}\leftrightarrow\xi_1^{}\,,\\
2x_2^{}&\textrm{for}&\xi_2^{}\leftrightarrow\xi_2^{}\,,\\
z &\textrm{for}&\phi_L^{}\leftrightarrow \phi_R^\ast\,,
\end{array}\nonumber
\end{eqnarray}
\vspace{-.6cm}
\begin{eqnarray}
\end{eqnarray}
which are constrained by
\begin{eqnarray}
x_1^{}+y+z=0\quad\textrm{and}\quad x_2^{}+y-z=0\,.
\end{eqnarray}
Clearly, the above assignment is consistent with the left-right
symmetry (\ref{lrsymmetry1}) and (\ref{lrsymmetry2}). The allowed
Yukawa interactions then should be
\begin{eqnarray}
\label{yukawa}\mathcal{L}_Y^{}&=&
-y_D^{}(\bar{q}_L^{}\widetilde{\phi}_L^{}D_R^{}+\bar{q}_R^{c}\widetilde{\phi}_R^{\ast}D_L^{c})-h_D^{}\xi_1^{}\overline{D}_L^{}D_R^{}\nonumber\\
&&-y_U^{}(\bar{q}_L^{}\phi_L^{}U_R^{}+\bar{q}_R^{c}\phi_R^{\ast}U_L^{c})-h_U^{}\xi_2^{}\overline{U}_L^{}U_R^{}\nonumber\\
&&-y_E^{}(\bar{l}_L^{}\widetilde{\phi}_L^{}E_R^{}+\bar{l}_R^{c}\widetilde{\phi}_R^{\ast}E_L^{c})-h_E^{}\xi_1^{}\overline{E}_L^{}E_R^{}\nonumber\\
&&-y_{N_R^{}}^{}(\bar{l}_L^{}\phi_L^{}N_{R}^{}+\bar{l}_R^{c}\phi_R^{\ast}N_{L}^{c})\nonumber\\
&&-y_{N_L^{}}^{}(\bar{l}_L^{}\phi_L^{}N_{L}^{c}+\bar{l}_R^{c}\phi_R^{\ast}N_{R}^{})\nonumber\\
&&-[f_N^{D}\overline{N}_L^{}N_R^{} +\frac{1}{2}f_N^{M}
 (\overline{N}_L^{} N_L^{c}+ \overline{N}_R^{c} N_R^{})]\xi_2^{}\nonumber\\
 &&+\textrm{H.c.}\,.
\end{eqnarray}
The fermion singlets can obtain their heavy masses after the global
symmetry breaking, i.e.
\begin{eqnarray}
\begin{array}{ccc}
M_D^{}=h_D^{}\langle\xi_1^{}\rangle\,,&M_U^{}=h_U^{}\langle\xi_2^{}\rangle\,,&
M_E^{}=h_E^{}\langle\xi_1^{}\rangle\,,\\
[2mm]
M_N^D=f_N^{D}\langle\xi_2^{}\rangle\,,&M_N^M=f_N^{M}\langle\xi_2^{}\rangle\,.
\end{array}
\end{eqnarray}

We emphasize that with the Yukawa couplings (\ref{yukawa}), the
strong CP phase will not vanish at tree level. The universal seesaw
models can provide a solution to the strong CP problem without an
axion \cite{begtsao1978} if the left-right symmetry is not the
charge-conjugation but the parity \cite{bm1990}. However, the parity
as the left-right symmetry is inconsistent with the global symmetry
(\ref{pqsymmetry}), which is essential for the mass generation of
the fermion singlets.

\emph{Universal seesaw}: By integrating out the charged fermion
singlets, $D_{L,R}^{}$, $U_{L,R}^{}$, and $E_{L,R}^{}$, we can have
the following dimension-5 operators,
\begin{eqnarray}
\mathcal{O}_5^{}&\supset&y_D^{}\frac{1}{M_D^{}}y_D^{T}\bar{q}_L^{}\widetilde{\phi}_L^{}\widetilde{\phi}_R^{\dagger}q_R^{}
+y_U^{}\frac{1}{M_U^{}}y_U^{T}\bar{q}_L^{}\phi_L^{}\phi_R^{\dagger}q_R^{}\nonumber\\
&&+y_E^{}\frac{1}{M_E^{}}y_E^{T}\bar{l}_L^{}\widetilde{\phi}_L^{}\widetilde{\phi}_R^{\dagger}l_R^{}
+\textrm{H.c.}\,.
\end{eqnarray}
After the left-right symmetry breaking, the right-handed charged
fermions can obtain their Yukawa couplings with the left-handed
fermion and Higgs doublets. Such Yukawa couplings, identified to
those in the SM, can generate the Dirac masses of the charged
fermions.

In the neutrino sector, the neutral fermion singlets $N_{L,R}^{}$
can form the Majorana or pseudo-Dirac fields, depending on the size
of $M_N^D$ and $M_N^M$. In the pseudo-Dirac case with $M_N^D\gg
M_N^M$, the induced dimension-5 operators are
\begin{eqnarray}
\mathcal{O}_5^{}&\supset&[y_{N_R^{}}^{}\frac{1}{M_N^{D}}y_{N_R^{}}^T+y_{N_L^{}}^{}\frac{1}{(M_N^{D})^T_{}}y_{N_L^{}}^T]
\bar{l}_L^{}\phi_L^{}\phi_R^{\dagger}l_R^{}
\nonumber\\
&&+\textrm{H.c.}\,,
\end{eqnarray}
which can account for the light Dirac neutrinos. In the Majorana
case with $M_N^M \gg M_N^D$, we obtain
\begin{eqnarray}
\mathcal{O}_5^{}&\supset&(y_{N_R^{}}^{}\frac{1}{M_N^{M}}y_{N_L^{}}^{T}+y_{N_L^{}}^{}\frac{1}{M_N^{M}}y_{N_R^{}}^{T})
\bar{l}_L^{}\phi_L^{}\phi_R^{\dagger}l_R^{}\nonumber\\
&&+\frac{1}{2}(y_{N_R^{}}^{}\frac{1}{M_N^{M}}y_{N_R^{}}^{T}+y_{N_L^{}}^{}\frac{1}{M_N^{M}}y_{N_L^{}}^{T})\nonumber\\
&&\times
(\bar{l}_L^{}\phi_L^{}\phi_L^{T}l_L^{c}+\bar{l}_R^{c}\phi_R^{\ast}\phi_R^{\dagger}l_R^{})+\textrm{H.c.}\,.
\end{eqnarray}
Clearly, the right-handed neutrinos can get a Majorana mass term
from the operators involving the right-handed doublets. Their Yukawa
couplings to the left-handed lepton and Higgs doublets can be
induced by the operators involving the left- and right-handed
doublets. This means we have realized the type-I seesaw
\cite{minkowski1977}. Furthermore, the operators involving the
left-handed doublets will give an additional contribution to the
neutrino masses, playing a role of the type-II seesaw \cite{mw1980}.

\emph{Peccei-Quinn symmetry}: We can classify our model into three
cases by choosing the quantum numbers of the fermion and Higgs
doublets, i.e.
\begin{enumerate}
\item[1.] The Higgs doublets are trivial under the global symmetry.
In this case, we have $x_1^{}=x_2^{}$ as $y\neq 0$ but $z=0$. Thus
we only need one complex scalar singlet, i.e. $\xi_1^{}=\xi_2^{}$.
\item[2.] The fermion doublet are trivial under the global symmetry.
In this case, we have $x_1^{}=-x_2^{}$ as $z\neq 0$ but $y=0$. There
is only one complex scalar singlet, i.e. $\xi_1^{}=\xi_2^{\ast}$.
\item[3.] The fermion and Higgs doublets are not trivial under the global symmetry.
In this case, the quantum numbers $x_1^{}$ and $x_2^{}$ keep
independent as $y\neq 0$ and $z\neq 0$. This means the existence of
two complex scalar singlets.
\end{enumerate}
In the following we will clarify the global symmetry in any cases is
identified with the PQ symmetry.

In the first case, the unique complex scalar singlet can be
described by
\begin{eqnarray}
\xi_1^{}=\xi_2^{}=\frac{1}{\sqrt{2}}(f+\rho)\exp\left(i\frac{a}{f}\right)\,,
\end{eqnarray}
with $a$ being the NGB. It is easy to derive the tree-level
couplings of the NGB to the colored fermions,
\begin{eqnarray}
\label{tree1} \mathcal{L}\supset
\frac{1}{2f}(\partial_\mu^{}a)\left(\sum_{Q}^{}\overline{Q}\gamma^\mu_{}\gamma_5^{}Q-\sum_{q}^{}\bar{q}\gamma^\mu_{}\gamma_5^{}q\right)\,,
\end{eqnarray}
where $Q$ denotes $D$ and $U$ while $q$ stands for the SM quarks.
The first term of the above Lagrangian associated with the
tree-level $Q-q$ mixing will result in
\begin{eqnarray}
\label{tree2} \mathcal{L}&\supset&
\frac{1}{2f}(\partial_\mu^{}a)\sum_{Q,q}^{}\bar{q}\gamma^\mu_{}\gamma_5^{}
q
\left(y_Q^{}\frac{\langle\phi_L^{}\rangle\langle\phi_R^{}\rangle}{M_Q^2}y_Q^T\right)\nonumber\\
&\sim&-\frac{1}{2f}(\partial_\mu^{}a)\sum_{Q,q}^{}\bar{q}\gamma^\mu_{}\gamma_5^{}q\left[\mathcal{O}\left(\frac{m_q^{}}{M_Q^{}}\right)\right]\,.
\end{eqnarray}
With the quark-gluon interactions, it will also induce
\begin{eqnarray}
\label{loop} \mathcal{L}\supset
\frac{1}{2f}(\partial_\mu^{}a)\sum_{Q,q}^{}\bar{q}\gamma^\mu_{}\gamma_5^{}q\left[\frac{\alpha_s^2}{\pi^2_{}}
\ln\left(\frac{M_Q^{}}{m_q^{}}\right)\right]
\end{eqnarray}
at two-loop order, as in the Kim-Shifman-Vainshtein-Zakharov
\cite{kim1979} (KSVZ) model. Clearly, the couplings of the NGB to
the SM quarks are dominated by
\begin{eqnarray}
\mathcal{L}\supset
-\frac{1}{2f}(\partial_\mu^{}a)\sum_{q}^{}\bar{q}\gamma^\mu_{}\gamma_5^{}q\,.
\end{eqnarray}
Because of the instanton interaction, the NGB can pick up a tiny
mass \cite{weinberg1978,bt1978},
\begin{eqnarray}
\label{amass1}
m_a^{2}=N^2_{}\frac{Z}{(1+Z)^2_{}}\frac{f_\pi^{2}}{f^2_{}}m_\pi^{2}\,,
\end{eqnarray}
where $N=3$ for three families of the SM quarks while $Z\simeq
m_u/m_d$.

In the second case, we also have one complex scalar singlet,
\begin{eqnarray}
\xi_1^{}=\xi_2^{\ast}=\frac{1}{\sqrt{2}}(f+\rho)\exp\left(i\frac{a}{f}\right)\,.
\end{eqnarray}
The couplings of the NGB to the colored fermions should be
\begin{eqnarray}
\mathcal{L}\supset
\frac{1}{2f}(\partial_\mu^{}a)\left(\overline{D}\gamma^\mu_{}\gamma_5^{}D
-\overline{U}\gamma^\mu_{}\gamma_5^{}U\right)\,.
\end{eqnarray}
Similar to Eqs. (\ref{tree2}) and (\ref{loop}), we compute the
induced couplings to the SM quarks,
\begin{eqnarray}
\mathcal{L}&\supset&
\frac{1}{2f}(\partial_\mu^{}a)\sum_{U,D,q}^{}\bar{q}\gamma^\mu_{}\gamma_5^{}q\left[\frac{\alpha_s^2}{\pi^2_{}}
\ln\left(\frac{M_D^{}}{M_U^{}}\right)\right.\nonumber\\
&&\left.+c_q^{}\mathcal{O}\left(\frac{m_q^{}}{M_Q^{}}\right)\right]\,,
\end{eqnarray}
where $c_q^{}=1$ for $u,c,t$ while $c_q^{}=-1$ for $d,s,b$. The mass
of the NGB is given by
\begin{eqnarray}
\label{amass2}
m_a^{2}&=&N^2_{}\frac{Z}{(1+Z)^2_{}}\frac{f_\pi^{2}}{f^2_{}}m_\pi^{2}\left[\frac{\alpha_s^2}{\pi^2_{}}
\ln\left(\frac{M_D^{2}}{M_U^{2}}\right)\right.\nonumber\\
&&\left.+\sum_{q}^{}c_q^{}\mathcal{O}\left(\frac{m_q^{}}{M_Q^{}}\right)\right]^2_{}\,.
\end{eqnarray}

In the third case, there are two complex scalar singlets,
\begin{eqnarray}
\xi_{1,2}^{}&=&\frac{1}{\sqrt{2}}(f_{1,2}^{}+\rho_{1,2}^{})\exp\left(i\frac{a_{1,2}^{}}{f_{1,2}^{}}\right)\,.
\end{eqnarray}
The couplings of the two NGBs to the colored fermions are given by
\begin{eqnarray}
\mathcal{L}&\supset&
\frac{1}{2f_1^{}}(\partial_\mu^{}a_1^{})\left(\sum_{D}^{}\overline{D}\gamma^\mu_{}\gamma_5^{}D
-\frac{x_1^{}+x_2^{}}{2x_1^{}}\sum_{q}^{}\bar{q}\gamma^\mu_{}\gamma_5^{}q\right)\nonumber\\
&&+\frac{1}{2f_2^{}}(\partial_\mu^{}a_2^{})\left(\sum_{U}^{}\overline{U}\gamma^\mu_{}\gamma_5^{}U
-\frac{x_1^{}+x_2^{}}{2x_2^{}}\sum_{q}^{}\bar{q}\gamma^\mu_{}\gamma_5^{}q\right)\,.\nonumber\\
&&
\end{eqnarray}
Clearly, this case is identified to the first or second case for
$x_1^{}=\pm x_2^{}$. So, let's consider $x_1^{}\neq \pm x_2^{}$. In
this case we can neglect the contributions from the colored fermion
singlets,
\begin{eqnarray}
\mathcal{L}&\supset&
-\frac{1}{2f_1^{}}(\partial_\mu^{}a_1^{})\sum_{q}^{}\bar{q}\gamma^\mu_{}\gamma_5^{}q\left(\frac{x_1^{}+x_2^{}}{2x_1^{}}\right)\nonumber\\
&&-\frac{1}{2f_2^{}}(\partial_\mu^{}a_2^{})\sum_{q}^{}\bar{q}\gamma^\mu_{}\gamma_5^{}q\left(\frac{x_1^{}+x_2^{}}{2x_2^{}}\right)\,.
\end{eqnarray}
The two NGBs thus get their tiny masses as below,
\begin{subequations}
\label{amass3}
\begin{eqnarray}
m_{a_1^{}}^{2}&=&N^2_{}\frac{Z}{(1+Z)^2_{}}\frac{f_\pi^{2}}{f_1^{2}}m_\pi^{2}\left(\frac{x_1^{}+x_2^{}}{2x_1^{}}\right)^2_{}\,,\\
m_{a_2^{}}^{2}&=&N^2_{}\frac{Z}{(1+Z)^2_{}}\frac{f_\pi^{2}}{f_2^{2}}m_\pi^{2}\left(\frac{x_1^{}+x_2^{}}{2x_2^{}}\right)^2_{}\,.
\end{eqnarray}
\end{subequations}

In the above three cases, the NGBs from the global symmetry breaking
all couple to the axial vector current of the SM quarks and hence
obtain the instanton-induced masses. So, the global symmetry is the
PQ symmetry as the pNGB acts as the axion. For convenience, we
express the axion mass as
\begin{eqnarray}
m_{a}^{}&=&\frac{\sqrt{Z}}{(1+Z)}\frac{f_\pi^{}}{f_{a}^{}}m_\pi^{}\simeq
6.2\,\mu\textrm{eV}\left(\frac{10^{12}_{}\,\textrm{GeV}}{f_a^{}}\right)\,,
\end{eqnarray}
where the axion decay constant $f_a^{}$ can be derived from Eq.
(\ref{amass1}), (\ref{amass2}) or (\ref{amass3}). The PQ symmetry
breaking scale should be high enough to fulfill the theoretical and
experimental constraints \cite{sikivie2008}. For example, the PQ
symmetry breaking may happen before inflation to avoid the
cosmological domain wall problem. With an appropriate PQ symmetry
breaking scale, the axion can act as the dark matter.

\emph{Leptogenesis}: After the left-right symmetry breaking, the
leptogenesis can be applied to explain the matter-antimatter
asymmetry in the universe. If the neutral fermion singlets and hence
the neutrinos form the pseudo-Dirac fields, the decays of the
neutral fermion singlets can produce a lepton asymmetry stored in
the left-handed lepton doublets and an equal but opposite lepton
asymmetry stored in the right-handed neutrinos. Since the effective
Yukawa interactions between the left- and right-handed neutrinos are
extremely weak, they will go into equilibrium at a very low
temperature where the sphaleron \cite{krs1985} action is not active.
Therefore, the sphaleron process can partially transfer the
left-handed lepton asymmetry to a baryon asymmetry. This
leptogenesis scenario with Dirac neutrinos is titled as
neutrinogenesis \cite{dlrw1999}. In the other case with the neutral
fermion singlets being the Majorana fields, we can have the
conventional leptogenesis with Majorana neutrinos.

\emph{Conclusion}: In this paper we connected the universal seesaw
scenario, where not only the neutral neutrinos but also the charged
fermions obtain their masses through the seesaw mechanism, to the PQ
symmetry for solving the strong CP problem. In our model, the
fermion singlets, including the color triplets for generating the
quark masses and the color singlets for generating the lepton
masses, have the Yukawa couplings to one complex scalar singlet or
two. The scalar singlet acquires a large VEV to spontaneously break
the global PQ symmetry. So, the fermion singlets can obtain their
heavy masses for the realization of the universal seesaw. The
colored fermion singlets mediate the PQ symmetry to the SM quarks at
tree and/or loop level so that the axion can pick up a tiny mass
through the color anomaly. We thus naturally related the PQ symmetry
to the neutrino mass-generation \cite{shin1987}. Our model also
accommodates the leptogenesis with Majorana or Dirac neutrinos.

ML is supported by the Sonderforschungsbereich TR 27 of the Deutsche
Forschungsgemeinschaft. PHG is supported by the Alexander von
Humboldt Foundation.

\end{document}